\documentclass[aps,draft,12pt]{revtex4-1}

\usepackage{graphicx}% Include figure files
\usepackage{dcolumn}% Align table columns on decimal point
\usepackage{bm}% bold math

\begin{document}

\preprint{}

\title{Particle physics processes in cosmology through an effective Minkowski space formulation and the limitations of the method}
%\author{}
%\email{}
%\affiliation{\\\\}

\author{Recai Erdem}
\email{recaierdem@iyte.edu.tr}
\author{Kemal G{\"{u}}ltekin}
\email{kemalgultekin@iyte.edu.tr}
\affiliation{Department of Physics \\
{\.{I}}zmir Institute of Technology\\
G{\"{u}}lbah{\c{c}}e, Urla 35430, {\.{I}}zmir, Turkey}

\date{\today}

\begin{abstract}
We introduce a method where particle physics processes in cosmology may be calculated by the usual perturbative flat space quantum field theory through an effective Minkowski space  description at small time intervals provided that the running of the effective particle masses are sufficiently slow. We discuss the necessary conditions for the applicability of this method and illustrate the method through a simple example. This method has the advantage of avoiding the effects of gravitational particle creation in the calculation of rates and cross sections i.e. giving directly the rates and the cross sections due to the scatterings or the decay processes.
\end{abstract}

%\pacs{}


\maketitle

\section{introduction}

The current standard model of cosmology $\Lambda$CDM \cite{Weinberg} although being quite successful at cosmological scales has some observational difficulties at smaller scales and some theoretical problems \cite{CDM,ccp}. One of the most popular and preferred alternatives to $\Lambda$CDM are those that employ scalar fields (such as quintessence) or vector or fermionic fields for dark energy or dark matter or for both \cite{DM-scalar-vector,DM-fermionic,DE-scalar-vector,DE-fermionic}. The scalar, vector, or fermionic fields are also employed for cosmic inflation \cite{DE-fermionic,inflation-vector}. For the study of the evolution of these fields at quantum level the use of quantum field theory in curved space is needed \cite{QFTC1,QFTC}.

In quantum field theory in curved space one needs to determine the mode function for a given cosmological background to calculate the cross section or the rate corresponding to a particle physics process. The mode function must be determined for each cosmological background separately, and this may be a complicated process in general that makes the results more complicated and less clear when compared to the usual Minkowski space perturbative quantum field theory calculations. Moreover, in the calculations in the context of quantum field theory in curved space, the rates and the cross sections involve contributions due to (spontaneous) gravitational particle production in addition to those due to interaction Lagrangians. This makes the problem of identifying the results of the calculation even more difficult. The contribution of gravitational production may be decreased by using adiabatic approximation \cite{Bunch,QFTC1}. In this way one may also make use of WKB type solutions for mode function, and assume the space be approximately Minkowskian for the modes with sufficiently short wavelengths. However, in general, the quantum field theory in curved space calculations may have significant departures from the results of S-matrix formulation of the usual Minkowski space quantum field theory even when one may employ adiabatic approximation \cite{Boyanovsky-new}. The source of the problem is that, although, in principle, Minkowski space is a good approximation for sufficiently small wavelengths, adiabatic approximation by itself is not sufficient to justify the use of the usual Minkowski space quantum field theory as a sufficiently good approximation. One needs a more concrete and more general condition to justify the use of approximate Minkowski space quantum field theory calculations in cosmology. In this study we introduce a new rigorous condition (in addition to adiabatic condition) that guarantees the use of Minkowski space quantum field theory calculations in cosmology as a good approximation.  Moreover, the studies in literature in that context \cite{Boyanovsky1,Boyanovsky2,Vilja} usually employ the simpler case of scalar fields conformally coupled to gravity or to the cases effectively equivalent to conformally coupled case. Even in that case the results of the calculations are rather complicated. In this paper, we consider scalars minimally coupled to gravity by using a method that employs an effective Minkowski space formulation. We also discuss the limitations of the applicability of this condition.

The effective Minkowski space formulation studied here was employed in the context of a specific model for formation of Bose-Einstein condensation in cosmology \cite{Erdem-Gultekin}. The aim of this paper is to provide a more general, formal and detailed treatment of this formulation i.e. of the effective Minkowski space formulation of quantum field theory (in the background of a Robertson-Walker metric). We employ the simple, yet elaborate enough Lagragian density of \cite{Erdem-Gultekin}, in particular, the $\chi\chi\,\rightarrow\,\phi\phi$ processes in this framework, to see the basic implications of the formulation in a simple setting. We show that the corresponding calculations may be done by using the tools of the usual perturbative Minkowski space quantum field theory in sufficiently small time intervals provided that the parameters of the model satisfy some conditions. We also show that there is a considerable parameter space that satisfies those conditions.

\section{the basic conditions for effective Minkowski space formulation}

Spacetime at cosmological scales is described by
\begin{equation}
ds^2\,=\,-dt^2\,+\,a^2(t)\left[\frac{dr^2}{1-kr^2}+r^2
(d\theta^2+\sin^2{\theta}d\phi^2)\right]\;. \label{e1}
\end{equation}
For simplicity we take $k=0$ which is in agreement with
observations \cite{PDG}.
We consider the following action in this space
\begin{eqnarray}
&S&
\,=\,\int\sqrt{-g}\;d^4x\,\frac{1}{2}\{-g^{\mu\nu}\left[\partial_\mu\phi\partial_\nu\phi\,+\,
\partial_\mu\chi\partial_\nu\chi\right]\,-\,m_\phi^2\phi^2\,-\,
m_\chi^2\chi^2\,-\,\mu\,\phi^2\chi\} \label{e2a} \\
&=&
\int\,d^3x\,d\eta\,\frac{1}{2}\{\,
\tilde{\phi}^{\prime\;2}-(\vec{\nabla}\tilde{\phi})^2
+\tilde{\chi}^{\prime\;2}-(\vec{\nabla}\tilde{\chi})^2
-\tilde{m}_\phi^2\tilde{\phi}^2-\,
\tilde{m}_\chi^2\tilde{\chi}^2
-\tilde{\mu}\,\tilde{\phi}^2\tilde{\chi}\,\}\;,
\label{e2b}
\end{eqnarray}
where prime denotes derivative with respect to conformal time $\eta$ \cite{QFTC} and
\begin{eqnarray}
&&d\eta\,=\,\frac{dt}{a(t)}~,
~\tilde{\phi}\,=\,a(\eta)\phi~,
~\tilde{\chi}\,=\,a(\eta)\chi~,~a^\prime\,=\,\frac{da}{d\eta}~,~\dot{a}\,=\,\frac{da}{dt}~,~a^{\prime\prime}\,=\,\frac{d^2a}{d\eta^2}
~,~\ddot{a}\,=\,\frac{d^2a}{dt^2} \nonumber \\
&&\tilde{m}_i^2\,=\,m_i^2a^2-\frac{a^{\prime\prime}}{a}\,=\,a^2\left(m_i^2-\frac{\ddot{a}}{a}-\frac{\dot{a}^2}{a^2}\right)~~i\,=\,\phi,\chi~,~~
\tilde{\mu}\,=\,a\,\mu\;.
\label{e2aa}
\end{eqnarray}

We will use the above action to study the essential aspects of a formulation (that we call effective Minkoski space formulation) that may be used to calculate cross sections and rates of particle physics processes in the background of a Robertson-Walker metric.
Note that, the set-up given above only serves as a simple and instructive toy model to study basics of this formulation while, in principle, this formulation is applicable to any model provided that the conditions introduced below are satisfied.

We assume that the rate of the decays and the scatterings are much larger than the Hubble parameter so that one may take the masses $\tilde{m}_\chi$, $\tilde{m}_\phi$ constant during times smaller than the average time between two individual processes (e.g. between two individual $\chi\chi\,\rightarrow\,\phi\phi$ processes) while the time dependence is observed only at cosmological scales.
This condition may be expressed as the variation in $\tilde{m}_\chi^2=a^2\left(m_\chi^2-\dot{H}-2H^2\right)$ and $\tilde{m}_\phi^2=a^2\left(m_\phi^2-\dot{H}-2H^2\right)$  during times $\Delta\,t$ smaller than the average time (e.g. $\frac{1}{n_\chi\beta\sigma\,v}$ for $\chi\chi\,\rightarrow\,\phi\phi$) between  two individual processes (where $\beta$ is the effective penetration depth of the incoming beam to the target, $n_\chi$ is the number density of the target particles, $\sigma$ is the cross section of the process, $v$ is the relative velocity of the incoming and the target particles) should be very small i.e.
\begin{equation}
\left|\frac{\Delta\,\tilde{m}^2}{\tilde{m}^2}\right|\,=\,
\left|\frac{\Delta\,t\,\left(\frac{d\,a^2\left(m^2-\dot{H}-2H^2\right)}{dt}\right)}{a^2\left(m^2-\dot{H}-2H^2\right)}\right|\,\ll\,1
~,~~\mbox{where}~~\Delta\,t\,\leq\,\frac{1}{n_\chi\beta\sigma\,v}~~\mbox{for}~~\chi\chi\,\rightarrow\,\phi\phi \;.\label{t1x}
\end{equation}
 Here $m$ denotes either of $m_\chi$ or $m_\phi$, $\dot{H}=\frac{dH}{dt}=\frac{\ddot{a}}{a}-H^2$. The upper bound in (\ref{t1x}) on $\Delta\,t$ is imposed because we take $\Delta\,t$ as the time interval where the particles may be considered as free particles.

If we let
\begin{equation}
H\,=\,\xi\,a^{-s}\;, \label{t9a3}
\end{equation}
which includes all simple interesting cases e.g. radiation, matter, stiff matter, cosmological constant dominated universes, then (\ref{t1x}) becomes
\begin{equation}
\left|\frac{2\Delta\,t\,a^2H\left[m_\chi^2-(s^2-3s+2)H^2\right]}{a^2\left(m_\chi^2-(2-s)H^2\right)}\right|
\,=\,\left|2H\Delta\,t\,\left(1-\frac{s(s-2)H^2}{m^2+(s-2)H^2}
\right)\right|\ll\,1 \;.\label{t1xxx}
\end{equation}

Another basic condition for applicability of an effective Minkowski space formulation
is that the variation of the effective coupling constant $\tilde{\mu}$
for a sufficiently small time interval $\Delta\,t$ in each interval $\eta_i\,<\,\eta\,<\,\eta_{i+1}$ i should be negligible i.e.
\begin{equation}
\frac{\Delta\,\tilde{\mu}}{\tilde{\mu}}\,=\,\frac{\frac{d\tilde{\mu}}{dt}\Delta\,t}{\tilde{\mu}}\,=\,H\,\Delta\,t\,\ll\,1~,~~\mbox{where}~~ \Delta\,t\,\leq\,\frac{1}{n_\chi\beta\sigma\,v} \;.\label{t4b}
\end{equation}

Note that the condition (\ref{t4b}) guarantees the condition (\ref{t1xxx}) provided that $\left|\left(1-\frac{s(s-2)H^2}{m^2+(s-2)H^2}
\right)\right|$ is not very large. $\left|\left(1-\frac{s(s-2)H^2}{m^2+(s-2)H^2}\right)\right|$ is very large only if $s$ is very large or if  $\frac{m_\chi^2}{H^2}$ is very close to $2-s$. Thus, (\ref{t4b}) is enough to guarantee (\ref{t1xxx}) essentially in all realistic cases. Moreover, (\ref{t4b}) implies that $\Delta\,t$ is much smaller than the Hubble time $\frac{1}{H}$.  Therefore, in the following we will simply suffice to impose the condition (\ref{t4b}) rather than (\ref{t1xxx}).

 \section{implication of the basic conditions and the adiabatic condition for mode functions}

The fields $\tilde{\phi}$ and $\tilde{\chi}$ may be expanded in their Fourier modes; for example,
\begin{equation}
\tilde{\chi}(\vec{r},\eta)\,=\,\frac{1}{\sqrt{2}}\int\,\frac{d^3\tilde{p}}{(2\pi)^\frac{3}{2}}\left[a_p^-\,v_p^*(\eta)e^{i\vec{\tilde{p}}.\vec{r}}
\,+\,a_p^+\,v_p(\eta)e^{-i\vec{\tilde{p}}.\vec{r}}\right]\;,
\label{e2aaa}
\end{equation}
where $\vec{r}=(\tilde{x}_1,\tilde{x}_2,\tilde{x}_3)$, $a_p$ are the expansion coefficients that are identified by annihilation operators after quantization, $v_p$ are the basic normalized solutions of the equation of motion (i.e. mode functions) of $\tilde{\chi}$. In the time interval between two particle physics processes (such as those in Figure \ref{fig:1}) one may consider $\tilde{\chi}$ particles to be free particles, so they satisfy
\begin{equation}
v_p^{\prime\prime}\,+\,\omega_p^2(\eta)\,v_p\,=\,0~~~\mbox{where}~~\omega_p\,=\,\sqrt{|\vec{\tilde{p}}|^2+\tilde{m}_\chi^2} ~~,~~~
v_p^\prime\,v_p^*-v_p\,v_p^{*\,\prime}\,=\,2i\;. \label{e2aab1}
\end{equation}
where $\tilde{}$'s in
(\ref{e2aab1}) refer to the action (\ref{e2b}) while the quantities without $\tilde{}$ will refer to (\ref{e2a}).

 We consider WKB-type solutions for (\ref{e2aab1}) \cite{QFTC}
\begin{equation}
v_p(\eta)\,=\,\frac{1}{\sqrt{W_p(\eta)}}exp{\left(i\int_{\eta_i}^\eta\,W_p(\eta)\,d\eta\right)}\;, \label{e2c1}
\end{equation}
where $W_p$, by (\ref{e2aab1}), satisfies
\begin{equation}
W_p^2\,=\,\omega_p^2-\frac{1}{2}\left[\frac{W_p^{\prime\prime}}{W_p}-\frac{3}{2}\left(\frac{W_p^\prime}{W_p}\right)^2\right]\;. \label{e2c2}
\end{equation}

If we let $|W_p^2-\omega_p^2|$ in (\ref{e2c2}) be small with respect to $\omega_p^2$, then we may take
\begin{equation}
^{(0)}W_p=\omega_p
\label{e2c3a}
\end{equation}
 as an approximation for $W_p$. Further, if we let the variation of $\omega_p^2$ with $\eta$ be small (i.e. if the contribution of gravitational particle production to the particle physics processes is small \cite{QFTC1}) then we may take
\begin{equation}
^{(2)}W_p=\omega_p\left(1-\frac{\omega_p^{\prime\prime}}{4\omega_p^3}+\frac{3\omega_p^{\prime\;2}}{8\omega_p^4}\right)~,~~\mbox{etc.}
\label{e2c3b}
\end{equation}
as higher order (better) approximations. These approximations are known as adiabatic approximations of order zero, order two etc. \cite{Bunch,QFTC1}.
 Note that $^{(0)}W_p\,=\,\omega_p$ = constant corresponds to Minkowski case, and $^{(2)}W_p$ is obtained by substituting  $^{(0)}W_p$ with time dependent $\omega_p$ on the right hand side of (\ref{e2c2}) and then Taylor expanding the square root for slowly varying $\omega_p$.

 The requirement of slow variation of $\omega_p^2$ with time introduced above may be expressed in a formal way as
 \begin{equation}
 \frac{\omega_p^{\prime}}{\omega_p^2}\,<\,1.
 \label{rev1}
 \end{equation}
 As $\frac{\omega_p^{\prime}}{\omega_p^2}$ gets smaller and smaller, the WKB approximate solutions in (\ref{e2c3b}) get closer and closer to the solution in (\ref{e2c3a}) which, in turn, gets closer and closer to the exact solution at time $\eta$. The above condition guarantees the more intuitive identification of adiabatic approximation as slow variation of energy during Hubble time since $\frac{\omega_p^{\prime}}{\omega_p^2}$$\,<\,$$\frac{\omega_p^{\prime}}{H\omega_p}$ for physically relevant modes inside horizon (i.e. for $\omega_p\,>\,H$). In the following, first we will show that the condition (\ref{t1x}) ensures adiabatic approximation for most of the physically relevant parameter space, then we will show that the condition (\ref{t1x}) makes $^{(0)}W_p$ a perfect approximation in each interval $\Delta\,t$.

 The requirement of slow variation of $\omega_p^2$  with time in (\ref{rev1}) is similar to the condition (\ref{t1x})
 \begin{equation}
\frac{\omega_p^{\prime}}{\omega_p^2}
\,=\,a\frac{\left(\frac{1}{\omega_p}\right)\frac{d\tilde{m}_\chi^2}{dt}}{2\omega_p^2}\,<\, a\frac{\left(\frac{1}{\omega_p}\right)\frac{d\tilde{m}_\chi^2}{dt}}{2\tilde{m}_\chi^2}
\,<\, a\frac{\left(\frac{1}{\tilde{m}_\chi}\right)\frac{d\tilde{m}_\chi^2}{dt}}{2\tilde{m}_\chi^2}\,=\,a\frac{\left(\frac{1}{(\Delta\,t)\tilde{m}_\chi}\right)\,(\Delta\,t)\frac{d\tilde{m}_\chi^2}{dt}}{2\tilde{m}_\chi^2}\;.
\label{rx1a}
\end{equation}
 We see that the main differences between (\ref{rev1}) and (\ref{t1x}) are the replacements of $\Delta\,t\,\leq\,\frac{1}{n_\chi\beta\sigma\,v}$ by $\frac{1}{\omega_p}$ or $\frac{1}{\tilde{m}_\chi}$ and of $\ll\,1$ by $<\,1$. We notice that (\ref{t1x}) ensures (\ref{rev1}) if $\frac{1}{\Delta\,t\;\tilde{m}_\chi}$ is not much greater than 1 i.e. if $\Delta\,t\;\tilde{m}_\chi$ is not much smaller than 1. To see the range of the applicability of this condition, we take $\Delta\,t\,=\,\frac{1}{n_\chi\beta\sigma\,v}$ and let $\frac{\tilde{m}_\chi}{n_\chi\beta\sigma\,v}\,>\,{\cal O}(1)$ i.e. $\tilde{m}_\chi\,>\,n_\chi\beta\sigma\,v$. This, for example, implies $\tilde{m}_\chi\,>\,\hbar\,n_\chi\beta\sigma\,v\,=\,2\times\,10^{-32}\;eV$ for $\hbar\simeq\,6.6\times\,10^{-16}\,eV.second$, $n_\chi= 10^8\,(meter)^{-3}$ (i.e. at the order of the photon number density in the universe), $\sigma=10^{-33}\,(meter)^2$ (i.e. in the order of the cross section of electromagnetic interactions), $\beta=1$, $v=c=3\times\,10^8\,meters/second$. Therefore, it is safe to say that (\ref{t1x}) ensures the applicability of (\ref{rev1}) for all reasonable values of parameters unless we do not take $\Delta\,t$ much smaller than $\frac{1}{n_\chi\beta\sigma\,v}$. In fact, it is evident from the above argument that in the case of $\frac{1}{n_\chi\beta\sigma\,v}\tilde{m}_\chi\,=\,{\cal O}(1)$, (\ref{t1x}) also ensures that
  \begin{equation}
\frac{\omega_p^{\prime}}{\omega_p^2}
\,\ll\,1\;.
\label{rx1aa}
\end{equation}
This implies that, in most of the cases, (\ref{t1x}) ensures the adiabatic approximation. Another, even more important, result is derived below.

By using the expression for $\omega_p$ in (\ref{e2aab1}) we find that
\begin{equation}
\left|\frac{\Delta\,\omega_p^2}{\omega_p^2}\right|\,=\,\left|\frac{\left(\frac{d\omega_p^2}{dt}\right)\,\Delta\,t}{\omega_p^2}\right|
\,=\,\left|\frac{\left(\frac{d\tilde{m}_\chi^2}{dt}\right)\,\Delta\,t}{\omega_p^2}\right|\,<\,
\left|\frac{\Delta\,\tilde{m}^2}{\tilde{m}^2}\right| .\label{t1xx}
\end{equation}
This implies that, if (\ref{t1x}) holds in a time interval $\Delta\,t$, then $\left|\frac{\Delta\,\omega_p^2}{\omega_p^2}\right|\,\ll\,1$ is ensured in the same interval i.e. (\ref{t1x}) ensures that $\omega_p$ may be taken to be almost constant in that interval. Note that this argument is true for any time interval $\Delta\,t$.
This, in turn, implies that
(\ref{e2c3a}) is a very good approximation to the exact solution and $\omega_p$ may be taken to be constant for each time interval provided that (\ref{t1x}) is satisfied in each interval $\Delta\,t\,\leq\,\frac{1}{n_\chi\beta\sigma\,v}$. In other words, the higher order approximations in (\ref{e2c3b}) converge to (\ref{e2c3a}) for reasonable values of the parameters provided that (\ref{t1x}) is satisfied. (A more explicit derivation of this result may be found in A.)  Hence we find that $W_p\,\simeq\,^{(0)}W_p=\omega_p=\mbox{constant}$ may be taken as a good approximation in each time interval.
Therefore in each time interval $\eta_i\,<\,\eta\,<\,\eta_{i+1}$ we may take $v_p\,\simeq\,v_p^{(i)}$ where
\begin{equation}
v_p^{(i)}(\eta)\,=\,\frac{1}{\sqrt{\omega_p^{(i)}}}exp{\left(i\omega_p^{(i)}(\eta-\eta_i)\right)}~~\mbox{where}~~\;
\omega_p^{(i)}=\omega_p(\eta_i)~,~~\eta_i\,<\,\eta\,<\,\eta_{i+1} \label{e2c4}
\end{equation}
Hence (\ref{e2aaa}) may be expressed as
\begin{eqnarray}
\tilde{\chi}^{(i)}(\vec{r},\eta)\,\simeq\,
\int\,
\frac{d^3\tilde{p}}{(2\pi)^\frac{3}{2}\sqrt{2\omega_p^{(i)}}}\left[a_p^{(i)\,-}\,
e^{i\left(\vec{\tilde{p}}.\vec{r}-\omega_p^{(i)}(\eta-\eta_i)\right)}
\,+\,a_p^{(i)\,+}
\,e^{i\left(-\vec{\tilde{p}}.\vec{r}+\omega_p^{(i)}(\eta-\eta_i)\right)}\right]\label{e2aaax} \\
\eta_i\,<\,\eta\,<\,\eta_{i+1}\;, \nonumber
\end{eqnarray}
where $^{(i)}$ refers to the $i$th time interval between the $i$th and $(i+1)$th processes and $\eta_{i+1}-\eta_i=\Delta\,t$.

To see the difference of this framework  with the standard framework for the study of quantum field theory in curved space, it may be useful to point out the similarities and the differences between the formulation in \cite{Boyanovsky2} and in this paper regarding their domains of the applicability and the forms of the expansion of $\tilde{\chi}$ in terms of mode functions. The condition (\ref{t1x}) may be expressed as
\begin{equation}
\frac{\Delta\,\tilde{m}_\chi}{\tilde{m}_\chi}
=\left(\frac{H\hbar}{E_k}\right)\left(\frac{E_k}{\tilde{m}_\chi}\right)\left(\frac{\Delta\tilde{m}_\chi}{H\hbar}\right)
\,\ll\,1\;,
\label{r1}
\end{equation}
where $E_k=\sqrt{\frac{|\vec{\tilde{p}}|^2}{a^2}+m_\chi^2}$. In \cite{Boyanovsky2} the condition $\frac{H\hbar}{E_k}\,\ll\,1$ is imposed, which is equivalent to (\ref{rx1aa}) in the case of conformally coupled scalars where $\tilde{m}=a\,m$. It is evident that the condition $\frac{H\hbar}{E_k}\,\ll\,1$ does not automatically imply (\ref{r1}) and vice versa. In most of the cases $\left(\frac{E_k}{\tilde{m}_\chi}\right)\,>\,1$, and it is quite possible that one may have $\left(\frac{E_k}{\tilde{m}_\chi}\right)\,\gg\,1$, $\left(\frac{\Delta\tilde{m}_\chi}{H\hbar}\right)\,\gg\,1$ while  $\frac{H\hbar}{E_k}\,\ll\,1$ or one may have $\left(\frac{E_k}{\tilde{m}_\chi}\right)\,\sim\,{\cal O}(1)$ while $\frac{\Delta\,\tilde{m}_\chi}{\tilde{m}_\chi}\,\ll\,1$ since $\left(\frac{E_k}{\tilde{m}_\chi}\right)$ or $\left(\frac{\Delta\tilde{m}_\chi}{H\hbar}\right)$ or both may be smaller than one. This difference results in different forms for the mode functions. The condition $\frac{H\hbar}{E_k}\,\ll\,1$
makes $a^{\prime\prime}\simeq\,0$ so that $\tilde{m}_\chi\simeq\,a\,m_\chi$. Therefore, in the case of \cite{Boyanovsky2} $\tilde{m}_\chi$ can not be taken to be constant, so the frequency $\omega_p$ in (\ref{e2aab1}) depends on $\eta$ in general in that case while in this study the condition (\ref{t1x}) (i.e. (\ref{r1})) guarantees that one may take $\tilde{m}_\chi$ constant in each $\Delta\,t$, so $\omega_p=\omega_p^{(i)}$ is constant in each interval $\eta_i\,<\,\eta\,<\,\eta_{i+1}$, hence the corresponding mode function is just that of the Minkowski space, namely, $e^{\pm\,i\omega_p^{(i)}(\eta-\eta_i)}$ in that interval.

The form of (\ref{e2b}) and the above analysis imply that in each interval $\eta_i\,<\,\eta\,<\,\eta_{i+1}$ we have an effective Minkowski space given by \cite{Parker}
\begin{equation}
d\tilde{s}^2\,=\,-d\eta^2\,+\,d\tilde{x}_1^2+d\tilde{x}_2^2+d\tilde{x}_3^2\;, \label{t1a}
\end{equation}
where the masses of the particles are time dependent, and $\tilde{x}_i$ are related to (\ref{e1}) by $d\tilde{x}_1^2+d\tilde{x}_2^2+d\tilde{x}_3^2$=$dr^2+r^2
(d\theta^2+\sin^2{\theta}d\phi^2)$. This together with the condition (\ref{t1x}) implies we may use the tools of the usual perturbative quantum field theory for calculation of the rates and cross sections in an effective Minkowski space given by (\ref{t1a})
for each $\Delta\,t$ in (\ref{t1x}). For example, for the process given in Figure \ref{fig:1}, one may take the masses be constant during $\Delta\,t$ and use the usual formulas for the rates and cross sections of the usual (Minkowski space) quantum field theory and then we take masses of the particles during the next $\Delta\,t$  be another constant and then calculate the rates and cross sections for that $\Delta\,t$, and so on. Eq.(\ref{t1x}), hence Eq.(\ref{rx1aa}) ensures that a possible contribution to the particle physics processes due to a change in effective masses and decay widths and gravitational particle production is small since the change in the effective mass of the particles is small \cite{QFTC,Bunch}.

\section{Additional conditions to be satisfied for applicability of the method}

Another constraint for this formulation is the cluster decomposition principle \cite{cluster-decomposition-1,cluster-decomposition-2,Weinberg2} which may be stated that as "the outcome of a scattering event, in which two or several particles come in close contact with each other is unaffected by the presence of any number of particles very far away, or differently stated, that several scattering events separated from each other by large distances are independent of each other" \cite{cluster-decomposition-1}. In fact, once the particles participating in the interactions may be expressed in terms of field expansions as in (\ref{e2aaa}) and then quantized, then the cluster-decomposition is guaranteed \cite{Weinberg2}. However such an expansion for individual particles may be inapplicable in some cases because the particles may tend to act collectively due to long-range correlations or due to long-range interactions. If the de Broglie wave-lengths of particles in a system of particles overlap then there may be long-range order (e.g. a Bose-Einstein condensation \cite{BEC}) so that the system of particles tend to move as a whole rather than acting independently. In that case an expansion of the form of (\ref{e2aaa}) can be done only for the particles that are not in the condensate system. Therefore we should impose the condition
\begin{equation}
\frac{h}{|\vec{\tilde{p}}|}\,\ll\,\tilde{n}_\chi^{-\frac{1}{3}} \label{t4}
\end{equation}
for the Broglie wavelength of the particles where $\vec{\tilde{p}}$ and $\tilde{n}_\chi$ are the momenta and the number density of the incoming particles in a scattering process. Therefore this method is not applicable in the case of very low momenta particles in the incoming beam. In the case of long-range forces also there may be collective behavior. This is characterized by the Compton wavelength of the particles mediating the interaction (since at non-relativistic case the corresponding interaction potential may be expressed as $V(r)\,\sim\,\frac{e^{-\alpha\,m_\chi\,r}}{r}$). In the case of the usual scattering processes the long-ranged character of electromagnetic interactions do not pose a significant problem since the atoms and molecules in the target are electrically neutral as a whole and the interactions take place inside each atom separately. However this an important issue for strong interactions inside nucleons where all densely distributed colored particles may be effected by the scattering, especially when the coupling constant may be large \cite{CDP-QCD}. Therefore we also impose that
\begin{equation}
\frac{h}{\tilde{m}_\chi\,c}\,\ll\,\tilde{n}_\phi^{-\frac{1}{3}} \;.\label{t4a}
\end{equation}
These conditions require that for the applicability of the method, the momenta and the density of incoming particles should not violate (\ref{t4}), and the mass of the intermediate particle $\tilde{m}_\chi$ and the density of the interacting particles $\tilde{n}_\phi$ should not violate (\ref{t4a}). These conditions become important at the extremely low momenta of incoming particles, and extremely low mass of intermediate particles (especially in the case of appreciably strong interactions) for extremely high number density of incoming particles.

After identifying the effective Minkowski spaces and the corresponding field expansions in (\ref{e2aaax}) the next step is to find the rates and the cross sections corresponding to some process. To this end, first we must identify the asymptotic states, the {\it in} and {\it out} states to be able to express the interaction Lagrangian in terms of these states and then calculate the corresponding scattering amplitudes \cite{QFT}. In the usual Minkowski space quantum field theory the {\it out} and {\it in} states are taken as the free particle states in the remote future and remote past i.e. taken as the states for ${\it t\,\rightarrow\,\infty}$ and ${\it t\,\rightarrow\,-\infty}$, respectively in text books. However, in practice these measurements are done by detectors that are located sufficiently far from the interaction region and the coming particles come from sufficiently far locations. In other words, in practice the {\it out} and {\it in} states are identified as those corresponding to ${\it t\,\rightarrow\,T}$ and ${\it t\,\rightarrow\,-T}$, respectively, where {\it T} is sufficiently large so that the {\it in} and {\it out} states may be identified as free states. In this study we take {\it T}$=\frac{1}{2}\Delta\,t$ as the half of the time between two interactions. By the conditions (\ref{t1xx}), (\ref{t4}), (\ref{t4a}) we guarantee the assumption of taking this time to be sufficiently large so that each incoming and outgoing particle can be treated to be a free particle, so one may use the tools of the usual perturbative Minkowski space quantum field theory. This point is especially important for applicability of S-matrix formulation of Minkowski space quantum field theory in this framework. Adiabatic condition (\ref{rev1}), even in the form of $\frac{\omega_p^{\prime}}{\omega_p^2}\,\ll\,1$, is not enough to ensure the applicability of S-matrix formulation of Minkowski space quantum field theory. In fact, $\frac{\omega_p^{\prime}}{\omega_p^2}\,\ll\,1$ only ensures (\ref{e2aaax}) in a time interval $\frac{1}{\omega_p}$ which should be much smaller than $T\,\sim\,\frac{1}{n_\chi\beta\sigma\,v}$ since, otherwise, one cannot identify free asymptotic states. Therefore, imposing (\ref{t1x}) or (\ref{t1xxx}) (or (\ref{t4b})) is essential for applicability of S-matrix formulation of Minkowski space quantum field theory in cosmology.

It is evident that (\ref{t1xxx}) can be satisfied in most of the times by taking $\Delta\,t$ sufficiently small. The only lower bound on $\Delta\,t$ comes from (\ref{t4}) and (\ref{t4a}) which are satisfied unless one takes momentum of the incoming particles extremely small and/or the masses of the intermediate particles and the number density of the final state particles violate (\ref{t4a}). In fact, in many cases, (\ref{t1xxx}) is satisfied even in the case where $\Delta\,t$ saturates the upper bound i.e. when $\Delta\,t=\frac{1}{n_\chi\beta\sigma\,v}$. In that case
(where $m=m_\chi$ and  $\Delta\,t=\frac{1}{n_\chi\beta\sigma\,v}$), (\ref{t1xxx}) becomes
\begin{equation}
\left|\frac{2a^2H\left[m_\chi^2-(s^2-3s+2)H^2\right]}{n_\chi\beta\sigma\,v\,a^2\left(m_\chi^2-(2-s)H^2\right)}\right|
\,=\,\left|\frac{2H}{n_\chi\beta\sigma\,v}\left(1-\frac{s(s-2)H^2}{m_\chi^2+(s-2)H^2}
\right)\right|\ll\,1 \;.\label{t1xxxa}
\end{equation}
Eq.(\ref{t1xxxa}) implies that either of $|\frac{2H}{n_\chi\beta\sigma\,v}|$, $|1-\frac{s(s-2)H^2}{m_\chi^2+(s-2)H^2}|$
is much smaller than one, and the other is at most at the order of one. In other words there are two extreme cases for (\ref{t1xxxa}) to be satisfied, ${\bf i}$- $|\left(1-\frac{s(s-2)H^2}{m_\chi^2+(s-2)H^2}\right)|$ has a value of at most, $|\left(1-\frac{s(s-2)H^2}{m_\chi^2+(s-2)H^2}\right)|\,\sim\,\mathcal{O}(1)$
provided that $s\simeq\,0$ or $s\simeq\,2$ or $\frac{m_\chi^2}{H^2}$ is not close to $2-s$. In that case it is enough to let $|\frac{2H}{n_\chi\beta\sigma\,v}|\,\ll\,1$ to satisfy (\ref{t1xxxa}). Note that $|\frac{H}{n_\chi\beta\sigma\,v}|\,<\,1$ should already be satisfied to enable the process to take place.
${\bf ii}$- $|\frac{2H}{n_\chi\beta\sigma\,v}|$ has a value of at most, $|\frac{2H}{n_\chi\beta\sigma\,v}|\,\sim\,\mathcal{O}(1)$. In that case it is enough to let $|\left(1-\frac{s(s-2)H^2}{m_\chi^2+(s-2)H^2}\right)|\,\ll\,1$ i.e. to set $\frac{m_\chi^2}{H^2}\,\simeq\,(s-1)(s-2)$ which also implies that $s\,>\,2$ or $s\,<\,1$ (in addition to $|\frac{m_\chi^2}{H^2}|\,\simeq\,|(s-1)(s-2)|$). Eq.(\ref{t1xxxa}) is satisfied for a considerable range of parameters. For example, the case i) may be realized in the radiation dominated era ($s=2$) well after its start (to make $|\frac{2H}{n_\chi\beta\sigma\,v}|\,\ll\,1$ applicable) independent of the value of $m_\chi$. The case i) is also satisfied for the current epoch of accelerated cosmic expansion (where $s\,\simeq\,0$) independent of the value of $m_\chi$ provided that $\frac{H_0}{n_0\beta\sigma_0\,v}\simeq\,\frac{H_0}{n_0\sigma_0\,v}\simeq\,\frac{10^{-26}m^{-1}}{n_0\sigma_0(v/c)}\,\ll\,1$.  The case ii) may be satisfied in a possible stiff matter dominated era ($s=3$) after inflation or in the current accelerated expansion era, $s\neq\,0\sim\,0$ provided that $m_\chi^2\sim\,2H_0^2$.

An important comment is in order at this point. In the case $m_i^2a^2\,<\,\frac{a^{\prime\prime}}{a}$ in (\ref{e2aa}) (which corresponds to $m_i^2\,<\,(2-s)H^2$ in the case of (\ref{t9a3})) the effective mass $\tilde{m}_i$ becomes tachyonic. However $\frac{a^{\prime\prime}}{a}$ gets sufficiently small by time so that the particle masses become real after some time for all physically relevant cases (e.g. as given in (\ref{t9a3})) except in the case where strictly $s=0$. Therefore this is not a true problem in general for the physically interesting cases because either the mass becomes real after a finite time for $s\,>\,0$ or it can not interact with other particles (so, making the tachyonic state harmless)for $s\,\leq\,0$ due to fast expansion rate. However a tachyonic state can not be dealt within this formulation because the would-be ground state (e.g. $\chi=0$, $\phi=0$) will not be the ground state anymore, making the perturbation expansion about the ground state inapplicable. In other words this formulation is not applicable to the case, $m_i^2a^2\,<\,\frac{a^{\prime\prime}}{a}$ (which corresponds to $m_i^2\,<\,(2-s)H^2$ in the case of (\ref{t9a3})). Therefore the case of $s\,\geq\,2$ (e.g. of radiation and stiff matter) is safe in this regard while, in the the case of $s\,<\,2$ (e.g. for cosmological constant and dust) $H^2$ should be sufficiently small compared to $m_\chi^2$ so that the problem of tachyons do not emerge. After combining this constraint with those discussed after (\ref{t1xxxa}) one notices that there is still a significant relevant available parameter space left. The conclusions obtained after (\ref{t1xxxa}) remain intact for radiation and stiff matter dominated eras, and the conclusions obtained for the current accelerated expansion era still hold provided that $m_\chi^2$ is not equal to $(2-s)\,H_0^2\hbar^2\,\sim\,10^{-66}$ eV$^2$.

\section{The allowed range of parameters}

Let us summarize what we have done up to this point: we have introduced three conditions, namely, (\ref{t1x}) (which in the case of (\ref{t9a3}) reduces to (\ref{t1xxx})), (\ref{t4}), (\ref{t4a}) for applicability of the method of approximating the Robertson-Walker space by Minkowski space in sufficiently small time intervals. One may also impose (\ref{t4b}) for a standard setup. Note that, (\ref{t4b}) guarantees (\ref{t1xxx}). Therefore, in physically relavant cases the essential conditions to be satisfied (for applicability of the effective Minkowski formulation introduced in this study) may be taken to be (\ref{t4}), (\ref{t4a}), (\ref{t4b}). Another constraint we had imposed is the exclusion of tachyonic states that is automatically guaranteed in a radiation dominated universe and for cosmological constant and matter dominated universe at current time is guaranteed for $m_\chi\,>\,\sim\,H_0\hbar\,\sim\,10^{-33}\,eV$  as we have mention in the preceding paragraph. We will assume that either of these conditions is satisfied i.e. no tachyonic states emerge.
Moreover, to guarantee the applicibality of (\ref{rx1}) it is useful to impose $\left(\frac{a}{\omega_p}\right)(\Delta\,t)^{-1}\sim\left(\frac{a\,n_\chi\beta\sigma\,v}{\omega_p}\right)\,<\,{\cal O}(1)$ as we have mentioned after Eq.(\ref{rx1}) although it is not a condition independent of (\ref{t1xxx}).

In other words, we have to impose the following conditions for the applicability of the effective Minkowski space formulation introduced in this study;
\begin{eqnarray}
&&\left(\frac{\tilde{m}_\chi\,c^2}{\hbar}\right)^3\,=\,\left(\frac{\tilde{m}_\chi\,c^2}{eV}\right)^3\left(\frac{eV}{\hbar}\right)^3\,\simeq\,1.3 \,\times\,10^{20}m^{-3}\left(\frac{\tilde{m}_\chi\,c^2}{eV}\right)^3\gg\,\tilde{n}_\phi\;, \label{t4axa} \\
&&\left(\frac{|\vec{\tilde{p}}|}{\hbar}\right)^3\,\simeq\,1.3 \,\times\,10^{20}m^{-3}\left(\frac{|\vec{\tilde{p}}|\,c}{eV}\right)^3\,\gg\,\tilde{n}_\chi \;,  \label{t4axb} \\
&&(H\,\Delta\,t)^{-1}\,\gg\,1~~\mbox{where}~~ \Delta\,t\,\leq\,\frac{1}{n_\chi\beta\sigma\,v}\;,  \label{t4axc} \\
&&\left(\frac{a}{\omega_p}\right)(\Delta\,t)^{-1}\sim\left(\frac{a\,n_\chi\beta\sigma\,v}{\omega_p}\right)\,<\,{\cal O}(1) \;,\label{t4axd}
\end{eqnarray}
 provided that $m_\chi\,>\,\sim\,H_0\hbar\,\sim\,10^{-33}\,eV$. Note that the condition (\ref{t4axa}) is imposed for the processes where the intermediate particle in the process is $\chi$ such as in the tree level
 $\phi\phi\,\rightarrow\,\phi\phi$. In the cases where the intermediate particle is $ \phi$, such as in the tree level
 $\chi\chi\,\rightarrow\,\phi\phi$ processes,  $\tilde{m}_\chi$ in (\ref{t4axa}) should be replaced by $\tilde{m}_\phi$. It is evident that, once (\ref{t4axa}) is satisfied, the corresponding expression where $\tilde{m}_\chi$ in (\ref{t4axa}) being replaced by $\tilde{m}_\phi$ will be automatically satisfied.

 Eq.(\ref{t4axa}) tells us that one can not use the scattering theory sufficiently well when the range of the forces are too long for a successful particle description for the individual particles in the system. This implies that, when the scattering of a system of charged particles is considered, the scattering cross section calculated in this formulation becomes less and less reliable when the mass of the intermediate particles goes to zero, especially when one considers high number densities of the scattering particles. However, this is not a big problem when one considers scatterers like atoms that are neutral up to very small distances. In fact, the bound (\ref{t4axa}) may be easily satisfied even for charged scatterers for reasonable choices of the number densities, and for sufficiently weak interactions as in \cite{Erdem-Gultekin}. For example, for $\tilde{m}_\chi\,c^2\,=\,1eV$ we have $\left(\frac{\tilde{m}_\chi\,c}{\hbar}\right)^3\,\sim\,10^{20}\,(meter)^{-3}$. Note that the number densities of nucleons, photons, and cold dark matter particles at present time are in the order of $\sim\,10^{-1}\,(meter)^{-3}$, $>\,10^8\,(meter)^{-3}$, and $\left(\frac{3\,\times\,10^{-3}eV}{\tilde{m}_\chi\,c^2}\right)$$\left(\frac{3\,\times\,10^{-3}eV}{\hbar\,c}\right)^3\,\simeq\,10^{13}\,\left(\frac{eV}{m_\chi\,c^2}\right)(meter)^{-3}$; respectively.

 The second condition (\ref{t4axb}) tell us that one can not use the scattering theory when the number density of the particles in the initial state becomes sufficiently high (or if the momentum is sufficiently low) so that Bose-Einstein correlation takes place and the system acts like a single quantity. For most of the parameter space this bound is not saturated. For example, we have $\left(\frac{|\vec{\tilde{p}}|}{\hbar}\right)^3\,\sim\,10^9\,(meter)^{-3}$ for $|\vec{\tilde{p}}|c\,\sim\,2\,\times\,10^{-4}\,eV$. This condition, for example, in \cite{Erdem-Gultekin} is guaranteed by taking the momenta of the particles in the initial state being high.

 The third condition (\ref{t4axc}) is the essential condition for validity of the effective Minkowski space formulation in this paper. It, in principle,  can be always satisfied by taking a smaller $\Delta\,t$ (provided that the interaction can be still taken to localized). However taking a smaller $\Delta\,t$ implies the excluding modes whose wavelengths greater than $\frac{|\vec{\tilde{p}}|}{\tilde{E}}\Delta\,t$. Therefore the optimum choice is to take $\Delta\,t\,=\,\frac{1}{n_\chi\beta\sigma\,v}$. In this case, one may put constraint on the possible values of $H$, $n_\chi$ and $\sigma$. For example, if one lets $H\,=\,H_0\,\sim\,2\,\times\,10^{-18}\,sec^{-1}$, $\sigma\,=\,10^{-40}\,(meter)^2$ (i.e. in the order of weak interaction cross sections), $\beta\sim\,1$, $v\sim\,c$, then one finds $n_\chi\,\gg\,\frac{H_0}{\beta\sigma\,v}\,\sim\,10^{14}\,(meter)^{-3}$ which may be only satisfied by cold dark matter particles of masses less than meV range. However, instead if one takes $\sigma\,=\,10^{-33}\,(meter)^2$ (i.e. in the order of electromagnetic interaction cross sections) and the same values of $H$, $\beta$, then one finds $n_\chi\,\gg\,\frac{H_0}{\beta\sigma\,v}\,\sim\,10^{7}\,(meter)^{-3}$ which is satisfied if one takes $n_\chi$ in the order of $10^{8}(meter)^{-3}$ i.e. in the order of the number density of nucleons in the universe. At earlier times this condition is usually satisfied more easily since $n\,\propto\,a^{-3}$ while $H\,\propto\,a^{-2}$ for a radiation dominated universe and $H\,\propto\,a^{-\frac{3}{2}}$ for a matter dominated universe.

 The last condition (\ref{t4axd}) may be satisfied provided that we consider the particles with momenta satisfying $\omega_p\,>\,a(t)\,n_\chi\beta\sigma\,v$. For example, for $n_\chi\,\sim\,10^{8}\,(meter)^{-3}$, $\beta\,\sim\,1$, $\sigma\,\sim\,10^{-32}\,(meter)^2$, $v\,\sim\,c$ at present we must have $\hbar\omega_p\,>\, 10^{-31}\,eV$ to be able to use this method.

 There is a considerable parameter space that satisfies all conditions in (\ref{t4axa}-\ref{t4axd}) simultaneously. For example, for $\Delta\,t\,=\,\frac{1}{n_\chi\beta\sigma\,v}$ the conditions (\ref{t4axb}) and (\ref{t4axc}) may be combined into
 \begin{equation}
 10^{20}m^{-3}\,\left(\frac{|\vec{\tilde{p}}|\,c}{eV}\right)^3\,\gg\,\tilde{n}_\chi\,\gg\,\frac{a^3\,H}{\beta\sigma\,v}\, \;,\label{t4axe}
 \end{equation}
 (where we have used $\tilde{n}_\chi=a^3\,n_\chi$) which, for example; for $|\vec{\tilde{p}}|\,c\,>\,1\,eV$, $\sigma\,=\,10^{-33}\,m^2$, $\beta\,\sim\,1$, $v\sim\,c$, $H\,=\,H_0\,\sim\,2\,\times\,10^{-18}\,sec^{-1}$ implies that  $10^{20}m^{-3}\,\gg\,\tilde{n}_\chi\,\gg\,10^7\,m^{-3}$. The equation (\ref{t4axa}) remains as an independent equation that restricts the number density of $\phi$ particles for a given $\tilde{m}_\chi$, or vice versa. The last equation (\ref{t4axd}) gives a lower bound for $\omega_p$ for a given interval of the values of $\tilde{n}_\chi$ as the one given in (\ref{t4axe}). For example, for the values given above it results in $\omega_p\hbar\,>\,10^{-29}\,eV$.

\section{Basic quantum field theory calculations in this framework}

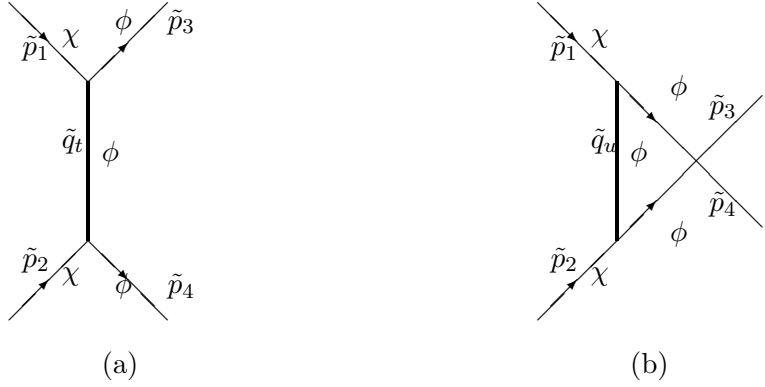
\begin{figure}[h]
\begin{picture}(200,150)(50,50)
\put (0,60){\vector(1,-1){15}}
\put (15,45){\line(1,-1){15}}
\put (30,30){\vector(1,1){15}}
\put (45,45){\line(1,1){15}}
\put (0,-60){\vector(1,1){15}}
\put (15,-45){\line(1,1){15}}
\put (30,-30){\vector(1,-1){15}}
\put (45,-45){\line(1,-1){15}}
\put(5,40){$\tilde{p}_1$}
\put(20,45){$\chi$}
\put(5,-40){$\tilde{p}_2$}
\put(20,-45){$\chi$}
\put(60,50){$\tilde{p}_3$}
\put(40,50){$\phi$}
\put(60,-50){$\tilde{p}_4$}
\put(40,-50){$\phi$}
{\bf {\linethickness{1pt}\put(30,-30){\line(0,1){60}}}}
\put(20,5){$\tilde{q}_t$}
\put(35,0){$\phi$}
\put(35,-80){(a)}
\hspace*{200pt}
\put (0,60){\vector(1,-1){15}}
\put (15,45){\line(1,-1){15}}
\put (30,30){\vector(1,-1){15}}
\put (45,15){\line(1,-1){40}}
\put (0,-60){\vector(1,1){15}}
\put (15,-45){\line(1,1){15}}
\put (30,-30){\vector(1,1){15}}
\put (45,-15){\line(1,1){40}}
\put(5,40){$\tilde{p}_1$}
\put(20,45){$\chi$}
\put(5,-40){$\tilde{p}_2$}
\put(20,-45){$\chi$}
\put(65,18){$\tilde{p}_3$}
\put(50,25){$\phi$}
\put(65,-19){$\tilde{p}_4$}
\put(50,-30){$\phi$}
{\bf {\linethickness{1pt}\put(30,-30){\line(0,1){60}}}}
\put(20,5){$\tilde{q}_u$}
\put(35,0){$\phi$}
\put(35,-80){(b)}
\end{picture}
\hspace{10pt}\\
\hspace{10pt}\\
\hspace{10pt}\\
\hspace{10pt}\\
\hspace{10pt}\\
\hspace{10pt}\\
\hspace{10pt}\\
\caption{The leading order Feynman diagrams that contribute to $\chi\chi\,\rightarrow\,\phi\phi$. Here
$\tilde{q}_t\,=\,\tilde{p}_1-\tilde{p}_3=\tilde{p}_4-\tilde{p}_2$,
$\tilde{q}_u\,=\,\tilde{p}_1-\tilde{p}_4\,=\tilde{p}_3-\tilde{p}_2$ are the 4-momenta carried in the internal lines. Note that the s-channel is forbidden by kinematics in Minkowski space. In a curved space, in general, one expects the s-channel to be opened due to the violation of energy/momentum conservation \cite{Boyanovsky2}. In this case where the space is taken to be nearly Minkowski in the intervals corresponding to average time for each process, one expects a small contribution to the scattering amplitudes due to the opened s-channel processes. However this contribution may be neglected since we take the space to be almost Minkowski in these time intervals by the condition (\ref{t1x}).  }
\label{fig:1}
\end{figure}

It is evident from (\ref{e2b}) that the cosmological time evolution of physical quantities may be considered mainly to be due to the time evolutions of the effective masses and the coupling constants. We have shown that, in each time interval $\Delta\,t\,=\,\frac{1}{n_\chi\beta\sigma\,v}$, one may consider the space approximately as a Minkowski space and  use the S-matrix formulation of the flat space quantum field theory provided that the condition (\ref{t1x}) (or (\ref{t4b})) and the conditions studied in Section IV are satisfied in that interval. Thus, one may let $\tilde{m}_\chi^2=\tilde{m}_\chi^2(\eta_r)$, $\tilde{m}_\phi^2=\tilde{m}_\phi^2(\eta_r)$,
$\tilde{\mu}=\tilde{\mu}(\eta_r)$ in the $r$'th interval $\eta_r\,<\,\eta\,<\,\eta_r\,+\Delta\,\eta$ (where $\Delta\,\eta$ is the conformal time interval corresponding to $\Delta\,t$), and then do the cross section or rate calculations by using the S-matrix formulation of the flat space space quantum field theory in that interval in the space described by (\ref{t1a}). After the calculation one may identify the $\tilde{m}_\chi^2(\eta_r)$, $\tilde{m}_\phi^2(\eta_r)$,  $\tilde{\mu}(\eta_r)$ terms in the result and replace them by $\tilde{m}_\chi^2(\eta)$, $\tilde{m}_\phi^2(\eta)$,  $\tilde{\mu}(\eta)$, and convert the momenta $\vec{\tilde{k}}$ and the energies $\tilde{E}$ for (\ref{t1a}) to the corresponding quantities for (\ref{e1}) by using $\vec{\tilde{k}}= a(t)\,\vec{k}$ (as discussed in Appendix B) to determine the final result with cosmic evolution. This method can be used for any particle physics process. In the following paragraphs, first we illustrate the method through a  simple example, namely, calculation of the total cross section for the Feynman diagrams in Figure \ref{fig:1}, then we present the general framework for this type of calculations.

 To illustrate how to apply this method we consider a simple example. We consider two types of particles $\chi$ and $\phi$ with significant cosmological number densities and making random collisions, and assume that all conditions for the applicability of this method hold. We also assume that the coupling constant $\mu$ is small enough so that one may use perturbative quantum field theory e.g.  we let $\frac{\tilde{\mu}}{\tilde{m}_\phi}\,\ll\,1$ for  Figure \ref{fig:1}. The corresponding cross section  in the effective Minkowski space is given by
\begin{equation}
\tilde{\sigma}\,=\,
\frac{(2\pi)^4}{4\sqrt{(\tilde{p}_1.\tilde{p}_2)^2-\tilde{m}_\chi^4}}\int\int\,\delta^{(4)}(\tilde{p}_1+\tilde{p}_2-\tilde{p}_3-\tilde{p}_4)\,|\tilde{M}|^2
\frac{d^3\vec{\tilde{p}}_3}{\tilde{E}_3}\frac{d^3\vec{\tilde{p}}_4}{\tilde{E}_4} \;,\label{t8}
\end{equation}
where $\tilde{m}_\chi$, $\tilde{m}_\phi$ are assumed to be constant during a time interval much smaller than the average collision time between two particles, $\tilde{M}$ corresponding to Figure \ref{fig:1} is given by
\begin{equation}
\tilde{M}\,=\,
\tilde{\mu}^2\left[
\frac{1}{(\tilde{p}_1-\tilde{p}_3)^2+\tilde{m}_\phi^2}
\,+\,\frac{1}{(\tilde{p}_1-\tilde{p}_4)^2+\tilde{m}_\phi^2}\right]\;.
\label{t1b}
\end{equation}
As we have mentioned before, the quantities with $\tilde{}$ correspond to the metric (\ref{t1a}) while those without $\tilde{}$ correspond to the metric (\ref{e1}). For example,
$E$ corresponds to (\ref{e1}) while $\tilde{E}$ corresponds to the same quantity for (\ref{t1a}) i.e.
\begin{equation}
E_i^2\,=\,a^2\left(m_i^2-\frac{\ddot{a}}{a}-\frac{\dot{a}^2}{a^2}\right)+a^2\vec{p}_i^2~~~\mbox{while}~~~
\tilde{E}_i^2\,=\,a^2m_i^2-\frac{a^{\prime\prime}}{a}+\vec{\tilde{p}}^2 \;.\label{s2}
\end{equation}
where we have used $\vec{\tilde{p}}=a\,\vec{p}$ (see Appendix B).
In the center of mass frame conservation of energy amounts to
\begin{equation}
\vec{\tilde{p}}^2+\tilde{m}_\chi^2=\vec{\tilde{k}}^2+\tilde{m}_\phi^2~~~\mbox{i.e.}~~~\vec{\tilde{p}}^2-\vec{\tilde{k}}^2=\tilde{m}_\phi^2 -\tilde{m}_\chi^2=a^2\left(m_\phi^2 -m_\chi^2\right) \;,\label{x}
\end{equation}
where $\vec{\tilde{k}}$ and $\vec{\tilde{p}}$ are $\vec{\tilde{p}}_3$ and $\vec{\tilde{p}}_1$ in the center of mass of the system, respectively. Note that $\vec{\tilde{p}}$, $\vec{\tilde{k}}$ do not depend on time, so (\ref{x}) is satisfied (provided that (\ref{t1x}) is satisfied), in general, only at the moment of each transition $\chi\chi\,\rightarrow\,\phi\phi$. It is evident from the right-hand side of (\ref{x}) that it may not hold at later times. Therefore one may use it in the evaluation of the cross sections and rates while it should not be used for studying the cosmological evolution of the relation between a particular $\vec{\tilde{p}}$, $\vec{\tilde{k}}$ pair (while it may be used for cosmological evolution of a generic $\vec{\tilde{k}}$ provided that cosmological evolution of a generic $\vec{\tilde{p}}$ is given). Another point is that if the time interval $\Delta\,t$ is taken extremely small to satisfy (\ref{t1x}) then the uncertainty introduced to $\tilde{E}=\sqrt{\vec{\tilde{p}}^2+\tilde{m}_\chi^2}$ by $\delta\,E=\frac{1}{\delta\,t}$ \cite{Boyanovsky2} may be comparable to or larger than $E$ itself, so making (\ref{x}) inapplicable. In fact, such a situation would also make (\ref{t8}) inapplicable since the energy conservation given in (\ref{x}) is also used in the temporal part of the four dimensional Dirac delta function in (\ref{t8}). However, we observe that $\delta\,E$ is small for most of the phenomenologically relevant set of parameters. This may be seen as follows. For $\delta\,t\sim\,\Delta\,t\sim\,\frac{1}{n_\chi\beta\sigma\,v}$, Eq.(\ref{t4axe}) may be reexpressed as
\begin{equation}
10^{20}m^{-3}\,\left(\frac{|\vec{\tilde{p}}|\,c}{eV}\right)^3\hbar\beta\sigma\,v\,\gg\,\delta\,E\,
\sim\,\frac{\hbar}{\Delta\,t}\,\sim\,\hbar\tilde{n}_\chi\,\beta\sigma\,v\,\gg\,\hbar\,a^3\,H. \label{ex1}
\end{equation}
For example, for the values after (\ref{t4axe}), namely, for $|\vec{\tilde{p}}|\,c\,>\,1\,eV$, $\sigma\,=\,10^{-33}\,m^2$, $\beta\,\sim\,1$, $v\sim\,c$, $H\,=\,H_0\,\sim\,10^{-18}\,sec^{-1}$, (\ref{ex1}) at present time ($a=1$) implies
$2\times\,10^{-20}\,eV$$\,\gg\,\delta\,E\,\sim\,\frac{\hbar}{\Delta\,t}\,\gg\,$$\,10^{-33}\,eV$. This, in turn, implies that $\frac{\delta\,E}{E}$ is negligible for these values of the parameters, provided that $E$ is greater than $2\times\,10^{-20}\,eV$. It is evident from (\ref{ex1}) that the upper bound on $\delta\,E$ may be increased by increasing the value of $|\vec{\tilde{p}}|\,\sigma\,\beta\,v$. For example, if $|\vec{\tilde{p}}|\,\sigma\,\beta\,v$ is increased $10^4$ times, then the upper bound on $\delta\,E$ changes as $2\times\,10^{-16}\,eV$$\,\gg\,\delta\,E$. In that case, taking $E$ be greater than $2\times\,10^{-16}\,eV$ ensures $\frac{\delta\,E}{E}$ be negligible while the conditions necessary for the applicability of effective Minkowski space remain intact. In other words, there is a considerable parameter space where one may use the Minkowski space quantum field theory formula (\ref{t8}) in each time interval $\Delta\,t$ for particle physics processes in cosmological backgrounds that satisfy the conditions discussed in the preceding section as a good approximation to an exact result that would follow from a standard quantum field theory in curved space calculation.

If we take $\tilde{m}_\phi\,\gg\,\tilde{m}_\chi$, $\vec{\tilde{p}}^2\,\gg\,\tilde{m}_\chi^2$, $\vec{\tilde{k}}^2\,\ll\,\tilde{m}_\phi^2$, then (\ref{t1b}) may be approximated by
\begin{equation}
\tilde{M}\,\simeq\,\left(\frac{\tilde{\mu}}{\tilde{m}_\phi}\right)^2\frac{1}{1-\frac{\vec{\tilde{k}}^2}{\tilde{m}_\phi^2}\cos^2{\theta}}\;.
\label{t1bx}
\end{equation}
In this case the cross section corresponding to Figure \ref{fig:1} is
\begin{equation}
\tilde{\sigma}\,
\simeq\,\left(\frac{\tilde{\mu}}{\tilde{m}_\phi}\right)^4
\frac{|\vec{\tilde{k}}|}{64\pi\,|\vec{\tilde{p}}|^2\tilde{m}_\phi} \;.\label{t8x}
\end{equation}
Next we relate $\tilde{\sigma}$ (which is expressed in terms of the coordinates in (\ref{t1a})) to the physically observed cross section $\sigma$ (which is expressed in terms of the coordinates in (\ref{e1})). We note that
\begin{equation}
\tilde{n}(\eta)\,=\,\frac{d\tilde{N}(\eta)}{d^3\tilde{x}}\,=\,\frac{dN(t)}{a^{-3}d^3x}
\,=\,a^3\,n(t)~~~\mbox{and}~~
a^4\left(\dot{n}_4+3H\,n_4\right)\,=\,\frac{d\,\tilde{n}_4(\eta)}{d\eta}\,=\,\beta\tilde{n}_1\tilde{n}_2\tilde{\sigma}\,\tilde{v} \label{t1sc}
\end{equation}
\begin{equation}
\dot{n}_4+3H\,n_4\,=\,\beta\,n_1n_2\sigma\,v \;.\label{t1sd}
\end{equation}
The equations (\ref{t1sc}) and (\ref{t1sd}) and $\frac{d|\vec{x}|}{dt}\,=\,v\,=\,\tilde{v}=\frac{d|\vec{\tilde{x}}|}{d\eta}$ together imply that $\sigma\,=\,a^2\,\tilde{\sigma}$. Therefore the cross section corresponding to Figure \ref{fig:1} is
\begin{equation}
\sigma\,
\simeq\,\frac{\mu^4}{\left(m_\phi^2-\frac{\ddot{a}}{a}-\frac{\dot{a}^2}{a^2}\right)^\frac{5}{2}}
\left(\frac{a^2|\vec{\tilde{k}}|}{64\pi\,|\vec{\tilde{p}}|^2}\right)\;. \label{t8xx}
\end{equation}
Hence once we know the scale factor one may determine the cosmological evolution of the cross section. Note that, even when $\frac{\ddot{a}}{a}+\frac{\dot{a}^2}{a^2}\,\sim\,0$, the physical cross section $\sigma$ has a dependence on the scale factor. For example, in a radiation dominated universe we have $\frac{\ddot{a}}{a}+\frac{\dot{a}^2}{a^2}\,=\,0$ while $a(t)$ varies with time considerably. Moreover, even when the variation of $a(t)$ in a time interval $\Delta\,t$ may be small, it may vary considerably during a Hubble time. To see the situation better, let $\frac{\ddot{a}}{a}+\frac{\dot{a}^2}{a^2}\,\sim\,0$ in (\ref{t8xx}). In that case, by (\ref{t4b}),  we get $\left|\frac{\Delta\,\sigma}{\sigma}\right|\,\ll\,1$ in a time interval $\Delta\,t\,=\,\frac{1}{n_\chi\beta\sigma\,v}$ as expected while $\left|\frac{\Delta\,\sigma}{\sigma}\right|$ in a Hubble time $\frac{1}{H}$ becomes of the order of one.

The relation $\sigma\,=\,a^2\,\tilde{\sigma}$ that is obtained above for Figure \ref{fig:1} may be seen more clearly by using a more formal consideration. The scattering amplitude from an in state (with $\,^1n$ particles with momenta ${\bf p}_1$, $\,^2n$ particles with momenta ${\bf p}_2$, etc.) to an out state  (with $\,^1m$ particles with momenta ${\bf k}_1$, $\,^2m$ particles with momenta ${\bf k}_2$, etc.) may be expressed as \cite{Birrell-Taylor,QFTC1}
\begin{eqnarray}
&&<\,out;\;^r\bar{m}_{k_r},\ldots,\,^2\bar{m}_{k_2},\,^1\bar{m}_{k_1}\,|\,^1n_{p_1},\,^2n_{p_2},\ldots\,^sn_{p_s}\;;in\,> \nonumber \\
~~&=&\sum_{h=0}^\infty\frac{1}{h!}\sum_{q_1\ldots\,q_h}\,<\,out;\;^r\bar{m}_{k_r},\ldots,\,^2\bar{m}_{k_2},\,^1\bar{m}_{k_1}\,|I_{q_1},I_{q_2},\ldots,I_{q_h};out> \nonumber \\
&&\times\,<out;I_{q_h},\ldots, I_{q_2},I_{q_1}|\,^1n_{p_1}\,^2n_{p_2},\ldots\,^sn_{p_s}\;;in\,> \;,\label{u1}
\end{eqnarray}
where the bars over the indices signify that they are expressed in terms of the creation operators of the out states $\bar{b}_{k_i}^\dagger$ (with $\bar{b}_{k_i}|0,out>\,=\,0$) while those without bar are expressed in terms of creation operators of the in states $a_{p_i}^\dagger$ (with $a_{p_i}|0,in>\,=\,0$); and $|I_{q_1},I_{q_2},\ldots,I_{q_h};out>$ denotes the complete set of all possible intermediate states in the final state that are expressed in terms of the creation operators of the in fields. Therefore, the terms $<\,out;\;^r\bar{m}_{k_r},\ldots,\,^2\bar{m}_{k_2},\,^1\bar{m}_{k_1}\,|I_{q_1},I_{q_2},\ldots,I_{q_h};out>$ do not contain information about the effect of the particle physics interactions, they only contain information about the effect of the curved space and are expressed in terms of Bogolyubov transformation matrices (see Appendix C). Note that the in and the out states in (\ref{u1}) are the same as the in and the out states in the effective Minkowski space formulation discussed in this study in the paragraph after Eq.(\ref{t4a}). The in and the out states in (\ref{u1}) are the usual in and the out states for $t\,\rightarrow\,-\infty$ and $t\,\rightarrow\,\infty$, respectively while the in and the out states in the paragraph after (\ref{t4a}) correspond to $t\,\rightarrow\,-L$ and $t\,\rightarrow\,L$, respectively, where $L$ is taken to be much larger than the range of the interactions so that may taken to be sufficiently good approximation to $t\,\rightarrow\,-\infty$ and $t\,\rightarrow\,\infty$. We remark that identification of in and out states in a standard quantum field theory in curved space analysis in general may be difficult. Although such an identification can be rigorously may be done for cosmological backgrounds that are localized in time, for general cosmological backgrounds such an identification is a highly non-trivial issue \cite{Wald}. On the other hand, the issue of the identification in and out states is not a problem in the present effective Minkowski space formulation since we take the space to be approximately  Minkowskian in each time interval $\Delta\,t=2T$ where we calculate the cross sections and rates, and we identify the in state as the state at $t\,\rightarrow\,-L$ and the out state as the state at $t\,\rightarrow\,L$. We use (\ref{u1}) and the formulas of the standard exact quantum field theory in curved space analysis in the following paragraphs to compare the result $\sigma\,=\,a^2\,\tilde{\sigma}$ obtained after (\ref{t1sd}) (by using the effective Minkowski space formulation) with the result of a standard exact quantum field theory in curved space analysis.

The transformation $\tilde{\phi}\,=\,a(\eta)\phi$ is an overall rescaling of the field both for in an out field operators, so it does not change the Bogolyubov transformation matrices. Therefore, the only effect of the field rescaling is through the second term that may be expressed as
\begin{eqnarray}
&&\frac{<out;I_{q_h},\ldots, I_{q_2},I_{q_1}|\,^1n_{p_1}\,^2n_{p_2},\ldots\,^sn_{p_s}\;;in\,>}{<out,0|0,in>}\,=\,i^{h+s}\,\prod_{i=1}^s\int\,d^4x_i\left[-g(x_i)\right]^{\frac{1}{2}} \nonumber \\
&&\times\,\prod_{j=1}^h\int\,d^4y_j\left[-g(y_j)\right]^{\frac{1}{2}}\,g_{p_i}(x_i)\, f_{q_j}^*(y_j)\;\left[-\nabla_{x_i\,\mu}\nabla_{x_i}^\mu+m^2+\xi\,R_{x_i}\right]
\left[-\nabla_{y_j\,\mu}\nabla_{y_j}^\mu+m^2+\xi\,R_{y_j}\right] \nonumber \\
&&\times\,\tau\left(y_h,\ldots,y_1,x_1,\ldots,x_s\right), \label{u2}
\end{eqnarray}
where $g_{p_i}(x_i)$, $f_{q_j}(y_j)$ are the mode functions defined in (\ref{xx1a}) and (\ref{xx1b}). We take $\xi=0$ in this study. The explicit form of the Green's function in (\ref{u2}) (for $s$ $\chi$ particles in the initial state and $h$ $\phi$ particles in the final state) is
\begin{equation}
\tau\left(y_1,\ldots,y_h,x_1,\ldots,x_s\right)\,=\,\frac{<out;0|T\left(\phi(y_1)\ldots\,\phi(y_h)\,\chi(x_1)\ldots\,\chi(x_s)\right)|0;in>}{<out,0|0,in>} \;. \label{u3}
\end{equation}
The corresponding expression in terms of $\tilde{\chi}$ and $\tilde{\phi}$ is given by
\begin{equation}
\tilde{\tau}\left(y_1,\ldots,y_h,x_1,\ldots,x_s\right)\,=\,\frac{<out;0|T\left(\tilde{\phi}(y_1)\ldots\,\tilde{\phi}(y_h)\,\tilde{\chi}(x_1)\ldots\,\tilde{\chi}(x_s)\right)|0;in>}{<out,0|0,in>} \;. \label{u3aa}
\end{equation}
It is evident that
$\tilde{\tau}$,
$\tau$ are related by $\tilde{\tau}\left(y_1,\ldots,y_h,x_1,\ldots,x_s\right)$
= $a^{h+s}\,\tau\left(y_1,\ldots,y_h,x_1,\ldots,x_s\right)$ since $\tilde{\chi}=a\,\chi$, $\tilde{\phi}=a\,\phi$ (see Appnedix D for a more formal and rigorous derivation).

Next, let us analyze the behaviour of the terms before the Green's function $\tau$ in (\ref{u2}) under the rescaling $\tilde{\chi}\,=\,a(\eta)\chi$, $\tilde{\phi}\,=\,a(\eta)\phi$.
By (\ref{e2aa}), (\ref{e2a}), (\ref{e2b}) we have $\sqrt{-\tilde{g}}\,d^4\tilde{x}\,$=$\,d\eta\,d^3x$, $\sqrt{-g}\,d^4x\,$=$\,a^3dt\,d^3x$, so $\sqrt{-\tilde{g}}\,d^4\tilde{x}\,$=$\,a^{-4}\sqrt{-g}\,d^4x$. The mode functions $g_{p_i}(x_i)$, $f_{q_j}(y_j)$ remain the same under the rescaling because  the equations of motion for $\tilde{\phi}$ and $\phi$ are related by
\begin{equation}
\left[-\tilde{\nabla}_{y_j\,\mu}\tilde{\nabla}_{y_j}^\mu+\tilde{m}^2\right]\tilde{\phi}\, = \,a^3(t)\,\left[-\nabla_{y_j\,\mu}\nabla_{y_j}^\mu+m^2\right]\phi\,
=\,0 \label{u3a}
\end{equation}
 so the corresponding solutions, namely mode functions, are the same. Moreover, (\ref{u3a}) implies that when each of $\left[-\nabla_{y_j\,\mu}\nabla_{y_j}^\mu+m^2\right]$ and  $\left[-\nabla_{x_i\,\mu}\nabla_{x_i}^\mu+m^2\right]$ in (\ref{u2}) acts to the corresponding $\chi$ and $\phi$ in (\ref{u3}), then each term results in a factor of $a^{-3}$ after the rescaling $\chi\,\rightarrow\,\tilde{\chi}$, $\phi\,\rightarrow\,\tilde{\phi}$. Hence, after combining these factors, one obtains
 \begin{eqnarray}
&&<\,out;\;^r\bar{m}_{k_r},\ldots,\,^2\bar{m}_{k_2},\,^1\bar{m}_{k_1}\,|\,^1n_{p_1},\,^2n_{p_2},\ldots\,^sn_{p_s}\;;in\,> \nonumber \\
~~&=&a^{h+s}\,<\,out;\;^r\bar{\tilde{m}}_{k_r},\ldots,\,^2\bar{\tilde{m}}_{k_2},\,^1\bar{\tilde{m}}_{k_1}\,|\,^1\tilde{n}_{p_1},\,^2\tilde{n}_{p_2},\ldots\,^s\tilde{n}_{p_s}\;;in\,>\;.
\label{u6a}
\end{eqnarray}
 i.e. it has the same scaling as the Green's function.

For Figure \ref{fig:1} we have $h=s=2$ in (\ref{u6a}), so, the amplitude after the scaling is $a^{-4}(t)$ times the one before rescaling. Note that the square of the absolute value of the amplitude in (\ref{u1}) when expressed in  in Minkowski space in terms of the rescaled fields is equal to $|\tilde{M}|^2$ in (\ref{t8}) up to the delta function in (\ref{t8}). Taking also the fact that $\vec{\tilde{p}}_i=a\,\vec{p}_i$, and $\tilde{E}_i=E_i$ by (\ref{s2}) into account we find that we should have $\tilde{\sigma}\,=\,a^{-2}\,\sigma$ which confirms the relation after Eq.(\ref{t1sd}) that is obtained in a less formal way.

\section{Conclusion}
 We have demonstrated that, if the variation of the effective masses defined by (\ref{e2aa}) are very small compared to the average collision time between the particles in the universe, then the spacetime may be taken to be approximately Minkowskian in these time intervals. Moreover, if the conditions of cluster decomposition principle are satisfied in these time intervals, and if the momenta of the incoming particles are sufficiently large to prevent a significant amount of Bose-Einstein correlation, and if the range of the interactions between the particles are small or are effectively short ranged, then one may use the tools of the usual Minkowski space quantum field theory to determine cross sections and the rates of the interactions between cosmic particles. We have shown that there is a considerable parameter space that satisfies these conditions. In the cases where this method is applicable it may be a simple tool to study the rates and the cross sections in cosmology.

\begin{acknowledgments}
We would like to thank Professor Daniel Boyanovsky for reading the draft version of this manuscript and for his valuable comments.

This paper is financially supported by {\it The Scientific and Technical Research Council of Turkey (T{\"{U}}BITAK)} under the project 117F296 in the context of the COST action {\it CA 16104 "GWverse"}
\end{acknowledgments}

\appendix

\section{Explicit derivation of the convergence of higher order approximations to the zeroth in the presence of (\ref{t1x}) for phenomenologically relevant set of parameters}

One may show that (\ref{t1x}) guarantees $\frac{\omega_p^{\prime\prime}}{4\omega_p^3}\,\simeq\,0$ and $\frac{3\omega_p^{\prime\;2}}{8\omega_p^4}\,\simeq\,0$  for reasonable values of the parameters, so one may take $W_p\,\simeq\,\omega_p$. This may be seen as follows: $\omega_p\,=\,\sqrt{|\vec{\tilde{p}}|^2+\tilde{m}_\chi^2}$ implies that
\begin{equation}
\frac{\omega_p^{\prime}}{\omega_p^2}\,=\,\frac{a}{\omega_p^2}\frac{d\omega_p}{dt}\,=\,\left(\frac{a}{2\omega_p^3}\right)\frac{d\tilde{m}_\chi^2}{dt}
\,=\,\left(\frac{a}{2\omega_p}\right)\left(\frac{\tilde{m}_\chi}{\omega_p}\right)^2(\Delta\,t)^{-1}
\left[\frac{\left(\frac{d\tilde{m}_\chi^2}{dt}\right)(\Delta\,t)}{\tilde{m}_\chi^2}\right]\,\simeq\,0\;,
\label{rx1}
\end{equation}
where $\Delta\,t\,\sim\,\left(\frac{1}{n_\chi\beta\sigma\,v}\right)$. By (\ref{t1x}), the last term in the square brackets in (\ref{rx1}) is much smaller than one, so in order to show that (\ref{rx1}) is almost zero the remaining terms in (\ref{rx1}) must not be much larger than one. For reasonable values of the parameters this is really the case. $\left(\frac{\tilde{m}_\chi}{\omega_p}\right)^2\,<\,1$, and $\left(\frac{a}{\omega_p}\right)(\Delta\,t)^{-1}\sim\left(\frac{a\,n_\chi\beta\sigma\,v}{\omega_p}\right)\,<\,{\cal O}(1)$ provided we take $\omega_p$ sufficiently large. This point will be discussed after (\ref{t4axd}).
Next we show that one may take $\frac{\omega_p^{\prime\prime}}{4\omega_p^3}\,\simeq\,0$ for reasonable values of the parameters. After using (\ref{t9a3}) and $\frac{d^2\omega_p^2}{dt^2}\,=\,\frac{d^2\tilde{m}_\chi^2}{dt^2}$, and  $\frac{d^2\omega_p^2}{dt^2}\,=\,2\left(\frac{d\omega_p}{dt}\right)^2+2\omega\frac{d^2\omega_p}{dt^2}$, and $\frac{d^2\tilde{m}_\chi^2}{dt^2}\,=\,2\left(\frac{d\tilde{m}_\chi}{dt}\right)^2+2\tilde{m}_\chi\frac{d^2\tilde{m}_\chi}{dt^2}$ we observe that
\begin{eqnarray}
&&\frac{d\tilde{m}_\chi}{dt}\,=\,\frac{1}{2\tilde{m}_\chi}\frac{d\tilde{m}_\chi^2}{dt}\,=\,H\tilde{m}_\chi\,+\,\frac{a^2s(2-s)H^3}{\tilde{m}_\chi}
\nonumber \\
&&\frac{d~^2\tilde{m}_\chi}{dt^2}\,=\,
(1-s)\,H^2\tilde{m}_\chi\,+\,\frac{a^2s(2-s)(2-3s)H^4}{\tilde{m}_\chi}\,-\,\frac{a^4s^2(2-s)^2H^6}{\tilde{m}_\chi^3}\label{rx2} \\
&&\mbox{so},~~~\frac{\omega_p^{\prime\prime}}{4\omega_p^3}\,=\,\frac{a}{4\omega_p^3}\frac{d\left(a\frac{d\omega_p}{dt}\right)}{dt}
\,=\,
\frac{a^2H\tilde{m}_\chi}{4\omega_p^4}\frac{d\tilde{m}_\chi}{dt}
\,+\,\frac{a^2}{4\omega_p^4}\left(1-\frac{\tilde{m}_\chi^2}{\omega_p^2}\right)\left(\frac{d\tilde{m}_\chi}{dt}\right)^2
\nonumber \\
&&+\,\frac{a^2\tilde{m}_\chi}{4\omega_p^4}\left[(1-s)\,H^2\tilde{m}_\chi\,+\,\frac{a^2s(2-s)(2-3s)H^4}{\tilde{m}_\chi}\,-\,\frac{a^4s^2(2-s)^2H^6}{\tilde{m}_\chi^3}\right]
\nonumber \\
&&=\,
\left(\frac{a}{\omega_p}(\Delta\,t)^{-1}\right)^2\left(\frac{\tilde{m}_\chi}{\omega_p}\right)^2
\{\frac{1}{8}H\Delta\,t\,
\left[\frac{\left(\frac{d\tilde{m}_\chi^2}{dt}\right)(\Delta\,t)}{\tilde{m}_\chi^2}\right]
\,+\,\frac{1}{16}
\left(1-\frac{\tilde{m}_\chi^2}{\omega_p^2}\right)\left[\frac{\left(\frac{d\tilde{m}_\chi^2}{dt}\right)(\Delta\,t)}{\tilde{m}_\chi^2}\right]^2
\nonumber \\
&&+\,\frac{1}{4}(H\Delta\,t)^2\left[
(1-s)\,\,+\,a^2s(2-s)(2-3s)\frac{H^2}{\tilde{m}_\chi^2}\,-\,a^4s^2(2-s)^2\left(\frac{H}{\tilde{m}_\chi}\right)^4\right]\}\,\simeq\,0\;,
\label{rx3}
\end{eqnarray}
provided that (\ref{t1x}) is satisfied.
 After using (\ref{rx1}) and (\ref{rx2}) one observes that (\ref{rx3}) too gives negligible contribution to (\ref{e2c3b}) for reasonable values of the parameters provided that either $H\Delta\,t$ and $\frac{H}{\tilde{m}_\chi}$ are not much larger than one. Hence we have shown that $W_p\,\simeq\,\omega_p$ in this case. In fact a similar result is obtained in \cite{Boyanovsky2} where the condition $\frac{H\hbar}{E_k}\,\ll\,1$ ($E_k=\sqrt{\frac{|\vec{\tilde{p}}|^2}{a^2}+m_\chi^2}$) is imposed. In that study $\omega_p$ is time dependent in general since the condition $\frac{H\hbar}{E_k}\,\ll\,1$ can not impose effective Minkowski spaces in small time intervals while the condition (\ref{t1x}) in this study makes
 $W_p\,\simeq\,^{(0)}W_p=\omega_p=\mbox{constant}$ in each time interval.

\section{Relation between the momenta with tilde and without tilde}

The relation between the momenta with tilde and without tilde in this paper is the same as the momenta in the geodesic equations for the metrics (\ref{e1}) and (\ref{t1a}) in a given time interval.
Therefore, the relation between the momenta with $\tilde{}$'s and those without $\tilde{}$'s is
\begin{equation}
|\vec{\tilde{p}}|\,=\,\tilde{m}_\chi\sqrt{\tilde{g}_{ij}\frac{d\tilde{x}^i}{|d\tilde{s}|}\frac{d\tilde{x}^j}{|d\tilde{s}|}}\,=\,
a\,m_\chi\sqrt{g_{ij}\frac{dx^i}{|ds|}\frac{dx^j}{|ds|}}\,=\,a |\vec{p}|\;. \label{t1y}
\end{equation}
 The equality in (\ref{t1y}) follows from extremization of the action for a free particle of mass $m$ i.e. $\int\,mds$ where $m=\tilde{m}_\chi$, $ds=d\tilde{s}$ for the metric (\ref{t1a}) and $m=m_\chi$, $ds=ds$ for the metric (\ref{e1}). In the case of (\ref{e1}) it reduces to the extremization of $\int\,ds$ since $m=m_\chi$ is constant. The resulting geodesic equations are $\frac{d\tilde{x}^i}{|d\tilde{s}|}\,=\,\frac{\sqrt{m_\chi\,v_i}}{\tilde{m}_\chi}$ and $\frac{dx^i}{|ds|}\,=\,\frac{\sqrt{v_i}}{a^2}$ where $v_i$ is some constant, so (\ref{t1y}) follows.
$|\vec{p}|\propto\,\frac{1}{a}$ is the physical momentum while $|\vec{\tilde{p}}|$ does not depend on redshift. We require that the variations in $\tilde{m}_\chi^2$ and $\tilde{m}_\phi^2$ are small (that are already ensured by (\ref{t1x})). This, in turn, implies that the variation in $\omega_p^2$ of (\ref{e2aab1}) with time is small since $\vec{\tilde{p}}$ does not depend on time.

\section{Derivation of the equations (\ref{u1}) and (\ref{u2})}

In general, the in and out field operators may be expressed as
\begin{eqnarray}
&&\chi_{in}(x)\,=\,\sum_n\left[a_n^{in\,-}f_n(x)\,+\,a_n^{in\,+}f_n^*(x)\right]\,=\,\sum_n\left[b_n^{in\,-}g_n(x)\,+\,b_n^{in\,+}g_n^*(x)\right] \label{xx1a} \\
&&\phi_{out}(x)\,=\,\sum_n\left[\bar{a}_n^{out\,-}f_n(x)\,+\,\bar{a}_n^{out\,+}f_n^*(x)\right]\,=\,\sum_n\left[\bar{b}_n^{out\,-}g_n(x)\,+\,\bar{b}_n^{out\,+}g_n^*(x)\right]\;, \label{xx1b}
\end{eqnarray}
where the creation, annihilation operators and the mode functions in the first and the second equations of (\ref{xx1a}) and (\ref{xx1b}) are related by Bogolyubov transformations. We have put an additional bar over the annihilation and the creation operators in (\ref{xx1b}) to simplify the notation in the following lines. The initial and the final vacuum states being identified by $a_n^{in\,-}|0,in>\,=\,0$, $\bar{b}_n^{out\,-}|0,out>\,=\,0$ for all $n$.
Consider the scattering amplitude from an in state $|^1n_{p_1}\,^2n_{p_2}.....\,^sn_{p_s}\;;in>$ to an out state $|^1\bar{m}_{k_1}\,^2\bar{m}_{k_2}.....\,^r\bar{m}_{k_r}\;;out>$, namely, \cite{Birrell-Taylor,QFTC1}
\begin{equation}
<out,\bar{m}^{(s)}|n^{(s)}>\,=\,<\,out;\;^r\bar{m}_{k_r}\,.....\,^2\bar{m}_{k_2}\,^1\bar{m}_{k_1}\,|\,^1n_{p_1}\,^2n_{p_2}.....\,^sn_{p_s}\;;in\,>\;,
\end{equation}
where $\,^nl_{k_j}$ stands for $\,l^n$ particles with momentum $k_j$, and
\begin{eqnarray}
&&|^1n_{p(1)}\,^2n_{p(2)}.....\,^sn_{p(s)}\;;in\,>\,=\,\left(^1n!\,^2n!\,......\,^sn!\right)^{-\frac{1}{2}}\,\left(a_{p(1)}^\dagger\right)^{\,^1n}\left(a_{p(2)}^\dagger\right)^{\,^2n},
\ldots,\left(a_{p(s)}^\dagger\right)^{\,^sn}\,|0,in>\;, \nonumber\\
&&|^1\bar{m}_{k(1)}\,^2\bar{m}_{k(2)},\ldots\,^r\bar{m}_{k(r)}\;;out>\,=\,\left(^1m!\,^2m!\ldots\,^rm!\right)^{-\frac{1}{2}}\left(\bar{b}_{k(1)}^\dagger\right)^{\,^1m}\left(\bar{b}_{k(2)}^\dagger\right)^{\,^2m},
\ldots,\left(\bar{b}_{k(r)}^\dagger\right)^{\,^rm}\,|0,out> \;.\nonumber \\
\label{x1b}
\end{eqnarray}
Then the corresponding amplitude $<out,\bar{m}^{(r)}|n^{(s)},in>$ may be found to be \cite{Birrell-Taylor}
\begin{eqnarray}
&&<out,\bar{m}^{(r)}|n^{(s)},in>
\,=\,\sum_\rho\sum_\sigma\,i^{\frac{(s-q)}{2}}i^{\frac{(r-q-w+k)}{2}}\{
\,O_{z\rho(1)}\ldots \,O_{z\rho(k)}\,
\,\alpha^{-1}_{\rho(k+1),\rho(k+1)}
\ldots\alpha^{-1}_{\rho(q),\rho(q)}\; \nonumber \\
~~~~&&\times\,\Lambda_{\rho(q+1),\rho(q+2)}\ldots
\Lambda_{\rho(s-1),\rho(s)}\,Q_{z\sigma(1)}\ldots \,Q_{z\sigma(w)}
\;V_{\sigma(w+1),\sigma(w+2)}\ldots\,V_{\sigma(r-q+k-1),\sigma(s-q+k-q)}
\,\nonumber \\
&&\times\,<out,0|\,T\left(\phi(z_{\sigma(1)})\ldots\,\phi(z_{\sigma(w)})\,\phi(z_{\rho(1)})\ldots\,\phi(z_{\rho(k)})\right)\,|0,in>
\}\;, \label{xx2}
\end{eqnarray}
 where the subscripts $z\rho(i)$ imply $z\rho(i)=z_{\rho(i)}$, and $\sum_\rho\sum_\sigma$ stands for the summations over $0\leq\,k\,\leq\,s$, $0\leq\,q\,\leq\,s-1$, $0\leq\,w\,\leq\,r$. Here the subindex $\rho$ stands for all possible combinations between $O_{z\rho()}$, $\alpha^{-1}_{\rho(),\rho()}$, $\Lambda_{\rho(),\rho()}$; and  the subindex $\sigma$ stands for all possible combinations between $O_{z\sigma()}$ and $V_{\sigma(),\sigma()}$, where
 \begin{eqnarray}
 &&O_{z(i)}= i\sum_p\,\alpha^{-1}_{ip}\,\int\,\sqrt{-g}\,d^4z_i\,g_p(z_i)\,\left(-\nabla_{z_i\,\mu}\nabla_{z_i}^\mu+m^2+\xi\,R_{z(i)}\right) \nonumber \\
 &&Q_{z(i)}= i\sum_p\,\alpha^{-1}_{pi}\,\int\,\sqrt{-g}\,d^4z_i\,f_p(z_i)\,\left(-\nabla_{z_i\,\mu}\nabla_{z_i}^\mu+m^2+\xi\,R_{z(i)}\right)  \nonumber\\
&& \Lambda_{ij}\,=\,-i\sum_p\beta_{pi}\alpha^{-1}_{jp}~, ~~V_{ij}\,=\,i\sum_p\beta_{jp}\alpha^{-1}_{pi} \nonumber
 \end{eqnarray}
with $m$ being the mass of $\phi$, $R$ is the curvature scalar, and $\xi=0$ in this case (i.e. for minimal coupling between matter and curvature). It is evident that (\ref{xx2}) has the form given in Eq.(\ref{u1}) \cite{QFTC1}, and after excluding the terms containing Bogolyubov coefficients $\alpha_{ij}$, $\beta_{ij}$ it corresponds to Eq.(\ref{u2}).

\section{Transformation of $\frac{<out;0|T\left(\phi(y_1)\ldots\,\phi(y_h)\,\chi(x_1)\ldots\,\chi(x_s)\right)|0;in>}{<out,0|0,in>}\,$
under $\chi\,\rightarrow\,\tilde{\chi}$, $\phi\,\rightarrow\,\tilde{\phi}$}

In this appendix we study the effect of $\chi\,\rightarrow\,\tilde{\chi}$, $\phi\,\rightarrow\,\tilde{\phi}$ on $\tau\left(y_1,\ldots,y_h,x_1,\ldots,x_s\right)$ in (\ref{u3}). We apply a careful path integral analysis since a naive analysis could be misleading. $\tau\left(y_1,\ldots,y_h,x_1,\ldots,x_s\right)$ in terms of functional derivatives may be expressed as \cite{QFTC1}
\begin{equation}
\tau\left(y_1,\ldots,y_h,x_1,\ldots,x_s\right)
\,=\,i^{-(h+s)}\left[
\frac{\delta^{h+s}\ln{Z[J]}}{\delta\,J_\phi(y_h)\ldots\,\delta\,J_\phi(y_1)\,\delta\,J_\chi(x_1)\ldots\,\delta\,J_\chi(x_s)}
\right]_{J=0}, \label{u4}
\end{equation}
The Green function generating functional for this study before the rescaling is
\begin{equation}
Z[J]\,=\,\int\,{\cal D}[\phi]\,{\cal D}[\chi]\;\exp{\left[i\,S\,+\,i\int\,\sqrt{-g}\,d^4x\,\left(J_\phi\,\phi+J_\chi\,\chi\right)\,\right]}\;.
\label{u5}
\end{equation}
 Here, \cite{path-integral1,path-integral2}
\begin{equation}
\int\,{\cal D}[\phi]\,=\,lim_{L\rightarrow\,\infty}lim_{N\rightarrow\,\infty}\,_{N\epsilon=L}\int\prod_m^N\,d\phi_m~,~~~\int\,{\cal D}[\chi]\,=\,lim_{L\rightarrow\,\infty}lim_{N\rightarrow\,\infty}\,_{N\epsilon=L}\int\prod_md\chi_m\;,
\label{apb1}
\end{equation}
where the 4-dimensional volume $L^4$ is divided into $N^4$ small cubes of sides $\epsilon$, the subindex $m$ denotes the $m$th cube.

The explicit form of $S$ in (\ref{u5}) is given in (\ref{e2a}), and
The $S$ before and after the rescaling are the same  by equality of (\ref{e2a}) and (\ref{e2b}) by construction. The $J_\phi\,\phi$ and $J_\chi\,\chi$ terms may be made to be invariant under rescaling by defining
\begin{equation}
\tilde{J}_\phi\,=\,a^3\,J_\phi\,,~~\tilde{J}_\chi\,=\,a^3\,J_\chi \;,\label{u5apx}
\end{equation}
where we have used $\tilde{\chi}=a\,\chi$, $\tilde{\phi}=a\,\phi$, $\sqrt{-\tilde{g}}\,d^4\tilde{x}\,$=$\,d\eta\,d^3x$ as mentioned before (\ref{u3a}).

 On the other hand, there could be some nontrivial effect of rescaling on (\ref{u5}) through the ${\cal D}[\phi]$ and ${\cal D}[\chi]$ terms. Another potential effect could be through the intermediate states since (\ref{u5}) may be also expressed as
 \begin{eqnarray}
Z[J]\,=&&\exp{\{i\int\,d^4x\;\mu\left(\frac{1}{i}\frac{\delta}{\delta\,J_{\phi}}\right)^2\left(\frac{1}{i}\frac{\delta}{\delta\,J_\chi}\right)\}} \nonumber \\
&&\times\,\int{\cal D}[\phi]{\cal D}[\chi]\,\exp{\left[i\,S_0+i\int\sqrt{-g}\,d^4x\,\left(J_\phi\phi+J_\chi\chi\right)\right]}\;.\label{u7}
\end{eqnarray}
 Hence a factor of $a^9$ will be introduced through the rescaling of $J_\phi$ and $J_\chi$ in the functional derivatives in (\ref{u7}) after using (\ref{u5apx}). However this rescaling does not effect (\ref{u4}) because of the presence of $\ln{Z}$ rather than $Z$. Because of the same reason, the effects of rescaling on ${\cal D}[\phi]$ and ${\cal D}[\chi]$  terms in (\ref{apb1}) cancel in (\ref{u4}).

In other words, the only effect of the rescaling $\chi\,\rightarrow\,\tilde{\chi}$, $\phi\,\rightarrow\,\tilde{\phi}$ on $\tau\left(y_1,\ldots,y_h,x_1,\ldots,x_s\right)$ is through the functional derivatives in (\ref{u4}) i.e. only through the rescaling of the fields $\phi$ and $\chi$ in (\ref{u3}) as expected.

\hspace{10pt}


\begin{thebibliography}{99}
\bibitem{Weinberg}
S. Weinberg, {\it Cosmology}
(Oxford Univ. Press, New York, 2008)
\bibitem{ccp} S. Weinberg, {\it The Cosmological Constant Problem}, {\it
Rev. Mod. Phys.} {\bf 61} 1 (1989); \\
S. Nobbenhuis, 2006 {\it Categorizing Different Approaches
to the Cosmological Constant Problem}, {\it Found. Phys.} {\bf 36}, 613 (2006); gr-qc/0411093
\bibitem{CDM}  D.H. Weinberg, et al., {\it Cold dark matter: Controversies on small scales}, {\it Proc. Nat. Acad.
Sci.} {\bf 112},   12249 (2015)\\
P. Bull, et al., {\it Beyond $\Lambda$CDM: Problems, solutions, and the road ahead}, {\it Physics of the Dark Universe}, {\bf 12}, 56 (2016)
\bibitem{DM-scalar-vector}
D. Azevedo, et.al., {\it Testing scalar versus vector dark matter}, {\it Phys. Rev. D} {\bf 99}, 015017 (2019), arXiv:1808.01598, and the references therein.
\bibitem{DM-fermionic}
A. Krut, C.R. Arg{\"u}lles, J. Rueda, R. Ruffini, {\it Glactic Constraints on Fermionic Dark Matter}, {\it Astron. Rep.} {\bf 62}, 898 (2018), and the references therein.
\bibitem{DE-scalar-vector}
R. Kase, S. Tsujikawa, {\it Dark energy in scalar-vector theories}, {\it JCAP} {\bf 1811}, 024 (2018), arXiv:1805.11919, and the references therein.
\bibitem{DE-fermionic}
G. Grams, R.C. Souza, G.M. Kremer, {\it Fermion field as inflaton, dark energy and dark matter}, {\it Class. Quant. Grav.} {\bf 31}, 185008 (2104), arXiv:1407.5481, and the references therein.
\bibitem{inflation-vector}
R. Emami, et.al, {\it Stable solutions of iflation driven by vector fields}, {\it JCAP} {\bf 1703}, 058 (2017), arXiv:1612.09581, and the references therein.
\bibitem{QFTC1}
N.D. Birrell and P.C.W. Davies, {\it Quantum fields in curved space}
(Cambridge Univ. Press, UK, 1994)
\bibitem{QFTC}
V. Mukhanov and S. Winitzki, {\it Introduction to Quantum Effects in Gravity}
(Cambridge Univ. Press, UK, 2007)
\bibitem{Bunch}
T.S. Bunch, {\it Adiabatic Regularization for scalar fields with arbitrary coupling to the scalar curvature},
{\it J. Phys. A: Math. Gen.} {\bf 13}, 1297 (1980); and the references therein.
\bibitem{Boyanovsky-new}
M. Rai and D. Boyanovsky, {Preprint},
(2020), arXiv:2012.10727.
\bibitem{Boyanovsky1}
D. Boyanovsky, {\it Condensates and quasiparticles in cosmology: Mass generation and decay widths},
{\it Phys. Rev. D} {\bf 85}, 123525
(2012), arXiv:1203.3903.
\bibitem{Boyanovsky2}
N. Herring, B. Pardo, D. Boyanovsky, A.R. Zentner, {\it Particle decay in post inflationary cosmology},
{\it Phys. Rev. D} {\bf 98}, 083503 (2018), arXiv:1808.02539
\bibitem{Vilja}
J. Lankinen, I. Vilja, {\it Decay of a massive Particle in a stiff-matter-dominated universe},
{\it Phys. Rev. D} {\bf 96}, 105026 (2017), arXiv:1709.07236;\\
J. Lankinen, I. Vilja, {\it Decaying Massive Particle in Matter and Radiation Dominated Eras},
{\it Phys. Rev. D} {\bf 97}, 065004 (2018), arXiv:1801.03757;\\
J. Lankinen, I. Vilja, {\it Particle decay in expanding Friedmann-Robertson-Walker universes},
{\it Phys. Rev. D} {\bf 98}, 045010 (2018), arXiv:1805.09620
\bibitem{Erdem-Gultekin}
R. Erdem, K. Gultekin, {\it A mechanism for formation of Bose-Einstein condensation in cosmology}, {\it JCAP} {\bf 1910}, 061 (2019), arXiv:1908.08784.
\bibitem{PDG} C. Patrignani et. al. (Particle Data Group),
{\it Review of Particle Physics (2016)}, {\it Chin. Phys. C} {\bf 40}, 100001 (2016)
\bibitem{Parker}
L. Parker, {\it Quantized Fields and Particle Creation in Expanding Universes. I}, {\it Phys. Rev.} {\bf 183}, 1057 (1969)
\bibitem{cluster-decomposition-1}
E.H. Wichmann, J.H. Crichton, {\it Cluster Decomposition Properties of S-Matrix}, {\it Phys. Rev.} {\bf 132}, 2788 (1963)
\bibitem{cluster-decomposition-2}
J.R. Taylor, {\it Cluster Decomposition of S-Matrix Elements}, {\it Phys. Rev} {\bf  142}, 1236 (1966)
\bibitem{Weinberg2}
S. Weinberg, {\it The Quantum Theory of Fields, Vol.I},
(Cambridge Univ. Press, New York, 1995)
\bibitem{BEC}
L. Pitaevskii, S. Stringari, {\it Bose-Einstein Condensation}, (Oxford Univ. Press, New York, 2003;\\
P.H. Chavanis, T. Harko, {\it Bose-Einstein condensate
relativistic stars}, {\it Phys. Rev. D} {\bf 86}, 064011 (2012),
arXiv:1108.3986
\bibitem{CDP-QCD}
P. Lowdon, {\it Conditions on the violation of the cluster decomposition property in QCD}, {\it J. Math. Phys.} {\bf 57}, 102302 (2016)
\bibitem{QFT}
C. Itzykson, J-B. Zuber, {\it Quantum Field Theory}, (Dover Pub., New York, 1980);\\
M.E. Peskin, D.V. Schroeder, {\it An Introduction to Quantum Field Theory}, (Addison-Wesley Pub., New York, 1995);\\
M. Maggiore, {\it A Modern Introduction to Quantum Field Theory}, (Oxford Univ. Press, New York, 2008);\\
M. Srednicki, {\it Quantum Field Theory}
(Cambridge Univ. Press, New York, 2007).
\bibitem{Wald} R.M. Wald, {\it Existance of the S-Matrix in Quantum Field Theory in Curved Space-Time}, {\it Annals. Phys.} {\bf 118}, 490 (1979).
\bibitem{Birrell-Taylor}
N.D. Birrell, J.C. Taylor, {\it Analysis of interacting quantum field theory in curved space-time}, {\it J. Math. Phys.} {\bf 21}, 1740 (1980).
\bibitem{path-integral1}
A. Das, {\it Field Theory: A Path Integral Approach}, (World Scientific, Singapore, 1993).
\bibitem{path-integral2}
L. Parker, D. Toms, {\it Quantum Field Theory in Curved Spacetime}, (Cambridge Univ. Press, UK, 2009).
\end{thebibliography}
\end{document}